\documentclass[10pt, twocolumn, aps, pre,showpacs]{revtex4}
\usepackage{natbib}
\usepackage{amsmath}
\usepackage{amsfonts}
\usepackage{amssymb}
\usepackage{ulem}
\usepackage{cancel}
\usepackage{graphicx}
\usepackage{subfigure}
\usepackage{braket}
\usepackage{placeins}
\usepackage{color}
\usepackage{latexsym}
\usepackage{pifont}
\usepackage{makecell}
\usepackage{wrapfig}
\begin{document}
\title{Hidden vortices and Feynman rule in Bose-Einstein condensates with density-dependent gauge potential}
\author{Ishfaq Ahmad Bhat, Bishwajyoti Dey}
\affiliation{Department of Physics, Savitribai Phule Pune University, Pune, Maharashtra, India, 411007}
\begin{abstract}
In this article, we numerically investigate the vortex nucleation in a Bose-Einstein 
condensate trapped in a double-well potential and subjected to a density-dependent 
gauge potential. A rotating Bose-Einstein condensate, when confined in a 
double-well potential, not only gives rise to visible vortices but also produces 
hidden vortices. We have empirically developed the Feynman’s rule for the number of 
vortices versus angular momentum in Bose-Einstein condensates in presence of the 
density-dependent gauge potentials. The variation of the average angular momentum 
with the number of vortices is also sensitive to the nature of the nonlinear rotation 
due to the density-dependent gauge potentials. The empirical result agrees well with 
the numerical simulations and the connection is verified by means of curve-fitting 
analysis. The modified Feynman rule is further confirmed for the BECs 
confined in harmonic and toroidal traps.  In addition, we show the nucleation of 
vortices in double-well and toroidally confined Bose-Einstein condensates solely through nonlinear rotations (without any trap rotation) arising through the density-dependent 
gauge potential.
\end{abstract}
\maketitle
\section{Introduction:}
\noindent
The initial realizations of Bose-Einstein condensates (BECs) employed harmonic traps (HT), allowing for a relatively simple and well-understood theoretical description of the system \cite{anderson_sc_1995, bradley_prl_1995, Fetter_rmp_2009, cooper_aip_2008}. Due to the technological developments in atomic, molecular and optical physics, the condensates are also obtained and subsequently studied in non-harmonic potentials, both experimentally and theoretically. Among the non-harmonic cases, the BECs confined in double well (DW) \cite{salasnich_pra_1999, albiez_prl_2005} and 
toroidal traps (TT) \cite{ryu_prl_2007, ramanathan_prl_2011, baharian_pra_2013} are studied in detail. The DW potential because of its richness in physics forms a significant model potential for studying the properties and the quantum phenomena in BECs on a mesoscopic scale \cite{smerzi_prl_1997, milburn_pra_1997, shin_prl_2004, shin_prl_2004atom, shin_prl_2005, saba_sr_2005, appmeier_thesis_2007}. Moreover, the static \cite{thomas_pra_2002, hofferberth_nat_2007,hall_prl_2007,jo_prl_2007} and the rotating \cite{hofferberth_nat_2006} DW potentials can be reliably realized with precision in present-day experiments. The rotating DW potentials, thus provide a platform to study the behaviour of the topological defects (quantized vortices) in these complex potentials. Similarly, the confinement of BECs within toroidal traps provides valuable insights into the intricate realm of quantum vortices \cite{ryu_prl_2007, aftalion_pra_2010}. 
\par 
Post successful realization in $1995$, BEC provided an amicable setting for studying the different aspects of the vortex physics \cite{Matthews_prl_1999, Madison_prl_2000, Madison_prl_2001, Hodby_prl_2001, Abosheer_sci_2001,  Recati_prl_2001, Sinha_prl_2001}. The quantized vortices within the BECs and superfluid $^{4}$He \cite{Donnelly_cup_1991} are chiefly nucleated by rotating the confining potential about a fixed axis \cite{Madison_prl_2000, Madison_prl_2001, Hodby_prl_2001, Abosheer_sci_2001}. The presence of vortices in BECs reveals their superfluid nature and has contributed to the understanding of nonlinear phenomena and phase coherence in quantum gases. The vortices in BECs are also known to play a key role in the understanding of quantum turbulence \cite{white_pnas_2014, tsatsos_pr_2016, mithun_pra_2021}. In harmonically confined BECs the vortices are visible within the density distribution and carry the angular momentum of the BEC. Additionally there exist the so-called \textit{ghost} vortices at the periphery of the condensate which do not carry any angular momentum. The total number of visible vortices, $N_{v}$ nucleated in an area $A$ reveals a linear relationship with the angular frequency of the trap, $\Omega$ as $2\pi \hbar N_{v}/m = 2 \Omega A$, where $m$ is the mass of the Bose-atom constituting the BEC \cite{Fetter_rmp_2009, cooper_aip_2008, Abosheer_sci_2001, haljan_prl_2001}. Feynman originally deduced this relation (Feynman's rule) in the context of superfluid Helium \cite{feynman_1955_chapter} and can be alternately expressed  in terms of  the average angular momentum of the rotating BEC as $\braket{L_{z}}/\hbar = N_{v}/2$.
\par 
However, in the BECs confined within the complex potentials, besides visible vortices there exist hidden vortices that are absent in the \textit{in situ} density profiles but have phase singularities. The phase profiles of the hidden vortices are similar to that of the visible vortices. The hidden vortices have no visible cores but carry angular momentum. In addition their core dimensions are determined by the barrier width rather than the healing length, which establishes the length scale for visible vortices. It is found that in the case of complex potentials, the celebrated Feynman's rule is satisfied only if the hidden vortices are also taken into account. In such cases, the Feynman's rule takes the form, $\braket{L_{z}}/\hbar = N_{t}/2$, where $N_{t} = N_{h} + N_{v}$ represents the sum total of hidden and visible vortices within the BEC \cite{Wen_pra_2010, Mithun_pra_2014, sabari_pla_2017}. Consequently, complex potentials and DW potentials, in particular, offer a platform for studying hidden vortices and thereby testing Feynman's rule in superfluids.
\par 
Charge-neutral BECs not only provide a flexible experimental platform to investigate various quantum phenomena, but they also provide a practical way to manifest artificial electromagnetism in experiments \cite{Dalibard_rmp_2011, Goldman_rpp_2014}. This achievement has been made possible through the creation of synthetic gauge potentials, accomplished either by rapidly rotating the condensate \cite{Matthews_prl_1999, Madison_prl_2000}, establishing optical connections between internal states of atoms \cite{Juzeliunas_prl_2004, Juzeliunas_pra_2006, Lin_nat_2009}, utilizing laser-assisted tunnelling \cite{ Aidelsburger_prl_2011,Miyake_prl_2013}, or employing Floquet engineering \cite{Parker_nat_2013}. The field of artificial gauge potentials is rapidly growing through experiments and theory. By inducing electromagnetism in BECs via experiments relying on Raman techniques \cite{Spielman_pra_2009}, new and intriguing phenomena have emerged, including orbital magnetism \cite{Lin_nat_2009, Lin_prl_2009}, exotic spin-orbit coupling \cite{Lin_nat_2011}, and spin-angular-momentum coupling with bosons \cite{Chen_prl_2018, Zhang_prl_2019} as well as fermions \cite{Wang_prl_2012,Cheuk_prl_2012}. However, the nature of these gauge potentials is typically static and lacks feedback between light and matter fields. On the other hand, in the case of dynamic gauge fields, there exists nonlinear feedback between the matter and gauge fields. Following the schemes \cite{Banerjee_prl_2012, Zohar_prl_2013, Tagliacozzo_nat_2013, Greschner_prl_2014, Dong_pra_2014, Ballantine_prl_2017} put forward for simulating dynamical gauge fields with ultracold atoms, experiments have successfully implemented the density-dependent gauge fields \cite{Clark_prl_2018, Gorg_nat_2019}. More recently, the density-dependent gauge potentials with a Raman coupled BEC within the context of a one-dimensional (1D) continuum was demonstrated experimentally \cite{frolian_nat_2022, chisholm_prr_2022,yao_nat_2022} The feedback mechanism between the light and matter fields in BECs with density-dependent gauge potentials is anticipated to enhance our understanding of atomic and nonlinear systems.
\par 
BECs featuring density-dependent gauge potentials, have already been investigated in the context of anyonic structures \cite{Keilmann_nat_2011}, chiral solitons \cite{Aglietti_prl_1996, Dingwall_njp_2018, Dingwall_pra_2019, Ishfaq_pre_2021}, collective excitations \cite{Edmonds_epl_2015}, and chaotic collective dynamics \cite{Chen_njp_2022}. The characteristic features of rotating two-dimensional harmonically confined density-dependent BECs have been studied numerically in \cite{Edmonds_pra_2020, Edmonds_pra_2021, Ishfaq_pre_2023}. The density-dependent gauge potential realizes an effective nonlinear rotation in BECs, resulting in the formation of vortex-lattices that do not exhibit the characteristic hexagonal symmetry of the Abrikosov vortices \cite{Campbell_prb_1979, Sato_pra_2007}. The observed patterns of vortices in BECs with density-dependent gauge potentials arise due to alterations in the Magnus force acting on the vortices and the repulsive interactions between them. Moreover, the critical trap rotation frequencies and the ellipticities also get modified, thereby depending on the \textit{s}-wave interaction strength. In a recent development \cite{Ishfaq_pre_2023} it has been revealed that vortex nucleation solely due to the density-dependent gauge potentials is not possible in the harmonically confined BECs. Further, the validity of Feynman's rule in BECs with density-dependent gauge potential also requires investigation.
\par 
In the present paper, we investigate the applicability of the established Feynman's rule 
in the BECs with density-dependent gauge potentials within the scope of direct numerical 
simulations. Our investigation encompasses a BEC confined within different potentials such 
as harmonic traps (HT), double-well potentials (DW), and toroidal traps (TT). Specifically, we 
examine the effects of the density-dependent gauge potentials on the dynamics of vortex 
nucleation in DW confined BECs while drawing comparisons across different confinements. 
We observe that BECs with density-dependent gauge potential do show deviations from 
the  Feynman's rule relating the number of vortices with the average angular momentum 
of the system. This is due to the fact that the  average angular momentum in a BEC with
density-dependent gauge potentials does not solely depend on the total number of 
vortices but also on the nature of the nonlinear rotation induced due to the density-dependent gauge potentials within the condensate. 
We also investigate the possibility of vortex nucleation solely due to the nonlinear
rotation in complex traps. We show in this study that vortex formation can be achieved in  BECs confined within the complex traps due to density-dependent gauge potentials and without any need for trap rotation. Furthermore, 
the manipulation of the strength and width of the potential barrier allows for the 
control over the occurrence of vortex nucleation.
\par
The subsequent material is structured as follows. In Section \ref{sec:model}, 
we present the Gross-Pitaevskii (GP) equation, which describes the dynamics of 
a trapped BEC while being subjected to a density-dependent gauge potential. 
This is followed by Sec. \ref{sec:numerics} where we discuss the results obtained from 
numerical investigations using the Crank-Nicolson method \cite{Muruganandam_cpc_2009, kumar_cpc_2019}. The work is finally summarized in Sec. \ref{sec:concl}.
%%%%%%%%%%%%%%%%%%%%%%%%%%%%%%%%%%%%%%%%%%%%%%%%%%%%%%%
\section{Model}\label{sec:model} 
 We consider a dilute BEC of $N$ two-level atoms coupled by a 
coherent light-matter interaction due to an incident laser 
field and described by the following mean-field Hamiltonian \cite{Dalibard_rmp_2011,Edmonds_pra_2020,Butera_jpb_2016}:
\begin{equation}\label{eq:ham}
    \hat{\mathcal{H}} = \left(\frac{\mathbf{\hat{p}}^2}{2m}+ V(\mathbf{r})\right)\otimes \check{\mathbb{I}} + \hat{\mathcal{H}}_{\text{int}} + \hat{\mathcal{U}}_{\text{lm}}
\end{equation}
where in the first term $\mathbf{\hat{p}}$ is the momentum operator, 
$V(\mathbf{r})$ is the trapping potential, and $\check{\mathbb{I}}$ is 
the $2\times 2$ unity matrix. The mean-field interactions, 
$\hat{\mathcal{H}}_{\text{int}} = (1/2)\text{diag}[\Delta_{1},\Delta_{2}]$
with $\Delta_{i}=g_{ii}|\Psi_{i}|^2 + g_{ij}|\Psi_{j}|^2$ and
$g_{ij} = 4\pi \hbar^2 a_{ij}/m$  
where $a_{ij}$ are the respective scattering lengths of the collisions
between atoms in internal states $i$ and $j$ ($i, j = 1, 2$). Further,
the light-matter interactions given by \cite{Goldman_rpp_2014,Butera_jpb_2016},
\begin{equation}\label{eq:lm}
    \hat{\mathcal{U}}_{\text{lm}} = \frac{\hbar \Omega_{r}}{2}
        \begin{pmatrix}
        \text{cos}\theta\left(\mathbf{r}\right)& e^{-i\phi\left(\mathbf{r}\right)}\text{sin} \theta\left(\mathbf{r}\right) \\
        e^{i\phi\left(\mathbf{r}\right)}\text{sin} \theta\left(\mathbf{r}\right) & -\text{cos} \theta\left(\mathbf{r}\right)
        \end{pmatrix}
\end{equation}
are parameterized in terms of Rabi frequency, $\Omega_{r}$,
mixing angle, $\theta\left(\mathbf{r}\right)$ and phase of the incident laser beam, $\phi\left(\mathbf{r}\right)$.
In a dilute BEC, when the Rabi-coupling energy is much larger than the 
mean-field energy shifts, the perturbative approach leads to
the following vector and scalar density-dependent potentials, $\mathbf{A}$ and 
$\mathit{W}$ respectively \cite{Edmonds_pra_2020,Butera_jpb_2016,Butera_pt_2017}:
\begin{equation}\label{eq:vpot}
    \mathbf{A} = -\frac{\hbar \theta^{2}}{4}\left(1-4\epsilon\right)\nabla\phi
\end{equation}
\begin{equation}\label{eq:spot}
    \mathit{W} = \frac{\hbar^{2}}{2}\left(\frac{\left(\nabla\theta\right)^2
    \left(1-4\epsilon\right)+\theta^2\left(1+4\epsilon\right)
    \left(\nabla\phi\right)^2}{4m} - \nabla\theta^2 \cdot\nabla\epsilon\right)
\end{equation}
where $\theta=\Omega_{r}/\Delta$ represents the ratio of Rabi frequency to 
laser detuning, and $\epsilon = \mathfrak{n}(g_{11}-g_{12})/4\hbar\Delta$ takes into 
account the collisional and coherent interactions. Within the limits of the 
adiabatic approximation wherein the coupled two-level atom is projected
onto a single dressed state, one arrives at the following mean-field 
Gross-Pitaevskii (GP) equation \cite{Edmonds_pra_2020,Edmonds_pra_2021, Ishfaq_pre_2023, Butera_jpb_2016,Butera_pt_2017,Butera_njp_2016}:
\begin{equation}\label{eq:gp3d}
\begin{split}
    i\hbar \frac{\partial\Psi}{\partial t} =& \left[ \frac{\left(\mathbf{\hat{p}} - \mathbf{A} \right)^2}{2m} +V(\mathbf{r})+\frac{\hbar\Omega_{r}}{2} + \delta + W + \mathbf{a}_{1}\cdot \mathbf{J}\right]\Psi \\
   & +\left[\mathfrak{n} \left(\frac{\partial W}{\partial \Psi^{*}} - \nabla \cdot \frac{\partial W}{\partial \nabla\Psi^{*}} \right) - \frac{\partial W}{\partial \nabla\Psi^{*}} \cdot \nabla \mathfrak{n}\right]
\end{split}
\end{equation}
where $\mathfrak{n} = |\Psi|^2$ is the density of the BEC in the projected state,  
$\delta = g_{11}\mathfrak{n} - \theta^2 \mathfrak{n}\left(g_{11}-g_{12} \right)/4 $ 
is the dressed mean-field interaction, and $\mathbf{a}_{1} = \theta \nabla\phi 
\left(g_{11}-g_{12}\right)/4\Delta$ is the strength of the coupling to 
the gauge field.  The current nonlinearity, 
\begin{equation}
    \mathbf{J} = \frac{\hbar}{2mi}\left[\Psi\left(\nabla + \frac{i}{\hbar}\mathbf{A}\right)\Psi^{*}-\Psi^{*}\left(\nabla -\frac{i}{\hbar}\mathbf{A}\right)\Psi\right]
\end{equation} in density-dependent BECs is 
a manifestation of the density-dependence of the geometric potentials 
\eqref{eq:vpot} and  \eqref{eq:spot}. 
By defining $\Omega_{\text{r}} = \kappa_{0}\mathit{r}$ 
where $\mathit{r}$ is the radial distance and 
$\phi= l \varphi$ where $l$ and $\varphi$ are respectively the angular 
momenta and polar angle of the laser beam \cite{Juzeliunas_pra_2005}, Eq. \eqref{eq:gp3d} 
after following the usual procedure of dimensional reduction 
\cite{Yunyi_mt_2005} results in the following two-dimensional (2D) 
Gross-Pitaevskii (GP) equation \cite{Edmonds_pra_2020,Edmonds_pra_2021, Ishfaq_pre_2023}:
\begin{equation}\label{eq:dmless_eq}
(i- \gamma) \frac{\partial\psi}{\partial t} = 
\bigg[-\frac{\nabla^2_{\boldsymbol{\rho}}}{2} +
V(\boldsymbol{\rho}) -\tilde{\Omega}_{n}(\boldsymbol{\rho},t) \hat{L}_{z} + 
\tilde{\text{g}} |\psi|^2 \bigg] \psi.
\end{equation} 
where
\begin{equation}\label{eq:nr}
    \tilde{\Omega}_{n}(\boldsymbol{\rho},t) = \Omega_{0}/\omega_{\perp} +
 \tilde{\mathit{C}}n(\boldsymbol{\rho},t)
\end{equation}
represents the density-dependent rotation experienced by the BEC because of the 
density-dependent gauge potentials. Here $\Omega_{0}$ is the trap rotation frequency 
in units of the transverse oscillation frequency, $\omega_{\perp}$ while 
$n(\boldsymbol{\rho},t)=|\psi(x,y,t)|^2$ is the 2D number density of the BEC in $xy$-plane.
In Eq. \eqref{eq:dmless_eq} which is a dimensionless GP equation, the operator 
$\nabla^2_{\boldsymbol{\rho}} =\frac{\partial^2}{\partial x^2} + 
\frac{\partial^2}{\partial y^2}$ while the interaction parameters,
$\tilde{\mathit{C}} = l \theta^2_{0} (g_{11}-g_{12})N/(\sqrt{2\pi}\sigma_{z}\hbar \Delta)$,
$ \tilde{\text{g}} =  g_{11}N m/\hbar^2 + \sqrt{2\pi}\sigma_{z}
(2 l^{2} -1)\tilde{\mathit{C}}/4l$. The coefficient $\theta_{0}=\kappa_{0}/\Delta$ and 
$\sigma_{z}$ represents the thickness of the BEC in the axial direction. 
In this dimensionless form the 2D wavefunction $\psi(\boldsymbol{\rho},t)$ is 
normalized to unity and $\hat{L}_{z} = -i\big(x \frac{\partial}{\partial y} - 
y\frac{\partial}{\partial x} \big)$ is the angular-momentum operator. 
Further, the BEC is assumed to be trapped either in an anisotropic HT \cite{Kasamatsu_pra_2003},
\begin{equation}\label{ht_pot}
    V(\boldsymbol{\rho}) = \frac{1}{2}\left[(1-\epsilon)x^2 + (1+\epsilon)y^2\right],
\end{equation}
where $\epsilon$ characterizes the anisotropy, 
or in the DW potential of form,
\begin{equation}\label{eq:dw_pot}
    V(\boldsymbol{\rho}) = \frac{\rho^{2}}{2} + V_{0}~\text{exp}(- x^{2}/2\sigma^2)
\end{equation}
with $V_{0}$ and $\sigma$ as the height and the width of the potential respectively. 
Moreover, in case of toroidal confinement, the trapping potential is expressed as 
\begin{equation}\label{eq:tt_pot}
    V(\boldsymbol{\rho}) = \frac{\rho^{2}}{2} + V_{0}~\text{exp}\left((- \beta x^{2} - y^{2}/\beta)/2\sigma^2\right)
\end{equation}
where $\beta$ is the anisotropy parameter. 
The parameter $\gamma = 0.03$ in Eq. \eqref{eq:dmless_eq} symbolizes dissipation within the system \cite{Tsubota_pra_2003, Kasamatsu_pra_2003, Mithun_pra_2014}. Its inclusion facilitates the faster convergence to an equilibrium state characterized by the presence of vortices without influencing the dynamics of the BEC \cite{Choi_pra_1998}. 
%%%%%%%%%%%%%%%%%%%%%%%%%%%%%%%%%%%%%%%%%%%%%%%%%%%%%
\section{Numerical Analysis}\label{sec:numerics}
The introduction of trap rotation, $\Omega_{0}$ has been employed both experimentally 
\cite{anderson_sc_1995, bradley_prl_1995,Madison_prl_2000, Madison_prl_2001} as well 
as theoretically \cite{Fetter_rmp_2009, cooper_aip_2008, kumar_cpc_2019, Kasamatsu_pra_2003} to evolve the anisotropically trapped BECs for vortex nucleation. 
The procedure is that a BEC is first prepared in a non-rotating ground state. In 
numerical simulations, this is achieved via the imaginary time propagation ($t \to -it$) of the GP equation \eqref{eq:dmless_eq} \cite{Muruganandam_cpc_2009, kumar_cpc_2019}. Following similar procedures and choosing $\Delta x = \Delta y = 0.08$ and $\Delta t = 0.001$ as space and time steps respectively, we first obtain the ground state of a BEC with $\Omega_{0}=\tilde{\mathit{C}}=0$ and confined either in a HT with $\epsilon=0.025$, TT with $\beta = 0.8$ or a DW potential with $\sigma = 0.5$. 
The vortex nucleation in the initially simulated ground state is then studied by evolving the GP equation \eqref{eq:dmless_eq} for different values of $\Omega_{0}$ and $\tilde{\mathit{C}}$. For each run, the initial condition $\psi^{0}_{j,k} = \sqrt{\text{max}(0,(\mu - V_{j,k})/g)} + \varepsilon ~\text{exp}(i \pi \mathcal{R}_{j,k})$, where the first term is the Thomas-Fermi wavefunction, $\varepsilon = 10^{-3}$ is the strength of the random perturbation to the wavefunction and $\mathcal{R}_{j,k}$ represents a pseudo-random matrix with elements from a uniform distribution in the range $[0, 1]$. This choice breaks any underlying symmetries within the system and prevents the simulation from getting stuck in any of the metastable states \cite{Edmonds_pra_2021}. 
\par
\begin{figure}[!htb]
\centering
\includegraphics[width=0.45\textwidth]{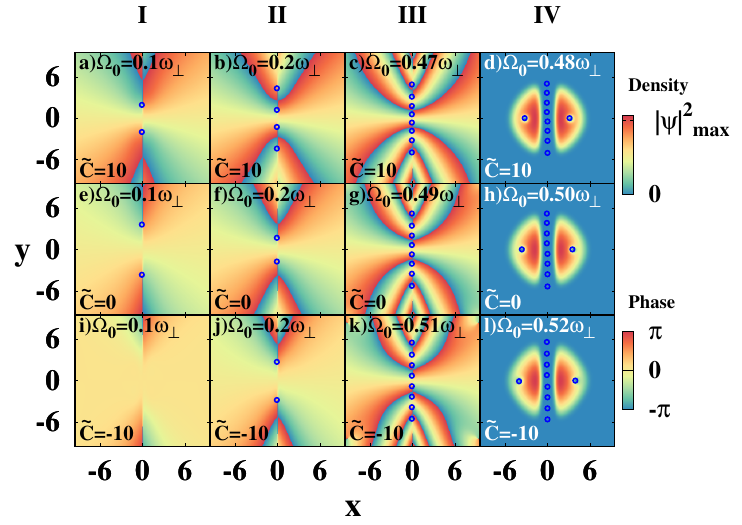}
\caption{Phase  (I-III) and Density (IV) profiles of the ground states in a BEC with  $\tilde{g} = 420$ and for mentioned strengths of nonlinear rotation, $\tilde{\mathit{C}}$ and trap rotations, $\Omega_{0}$. The height of the potential barriers is maintained at $V_{0}=40$ while its width, $\sigma = 0.5$. The blue circles mark the positions of the vortices in the phase and density profiles. 
}\label{fig:om_cr_den}
\end{figure}
\begin{figure}[!b]
\centering
\includegraphics[width=0.45\textwidth]{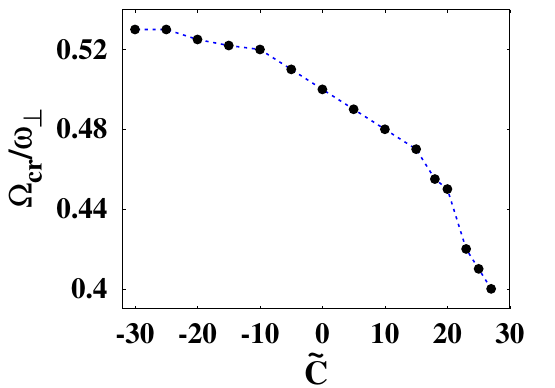}
\caption{Variation of the critical frequency ($\Omega_{\text{cr}}$) with the strength of the nonlinear rotation, $\tilde{\mathit{C}}$ in a DW confined BEC with  $\tilde{g} = 420$. The height of the potential barriers is maintained at $V_{0}=40$ while its width, $\sigma = 0.5$.
}\label{fig:om_cr_vs_nr}
\end{figure}
\begin{figure}[!htb]
\centering
\includegraphics[width=0.45\textwidth]{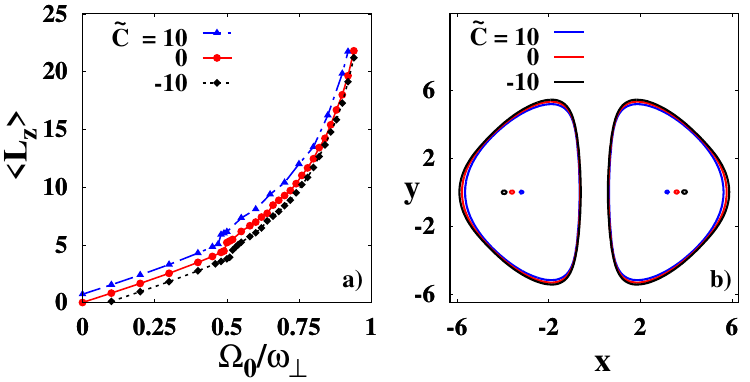}
\caption{a) Variation of the average angular momentum, $\braket{L_{z}}$ with trap rotation frequency, $\Omega_{0}$ in a rotating BEC with  $\tilde{g} = 420$ and for different strengths of nonlinear rotation, $\tilde{\mathit{C}}$. 
b) Contour plot of the vortex-ground states obtained at the respective critical frequencies. The small point like contours represent the positions of the visible vortices on either side of the central barrier of the DW potential. 
}\label{fig:lz_vs_om}
\end{figure}

The effects of density-dependent gauge potentials on the dynamics of vortex nucleation in a harmonically confined BEC is presented in \cite{Ishfaq_pre_2023}.
The process of vortex formation in a rotating DW potential differs greatly from that in a harmonic potential. In the case of harmonically confined BECs, only ghost vortices exist at the periphery of the cloud, and the BEC lacks substantial angular momentum when the trap rotation is less than the critical frequency ($\Omega_{\text{cr}}$). Once the trap rotation frequency exceeds the critical value, visible vortices emerge, accompanied by a significant increase in the angular momentum of the BEC. However, for a DW confined BEC, the process of vortex formation starts with the creation of a pair of hidden vortices. By following the figures of any row in Fig. \ref{fig:om_cr_den}, it is clear that a pair of hidden vortices emerge initially at the edges of the potential barrier which subsequently move towards the centre. The increase in trap rotation sees additional pairs of hidden vortices emerging along the barrier. The ghost vortices appear only after several pairs of hidden vortices have formed. 
%%%%%%%%%%%%%%%%%%%%%%%% FIG 
%\FloatBarrier 
A similar effect due to the nonlinear rotation, $\tilde{\mathit{C}}|\psi|^2$ induced by the density-dependent gauge potentials is expected. By examining the figures in columns I-III respectively, it is clear that the number of hidden vortices decreases and their separation increases as the value of $\tilde{\mathit{C}}$ decreases. These modifications also alter the critical frequency ($\Omega_{\text{cr}}$) as shown in Fig. \ref{fig:om_cr_vs_nr}, whereby a pair of visible vortices materializes in the ground state of the BEC. The critical frequency for vortex nucleation is observed to decrease with the increasing strength of nonlinear rotation, similar to the behavior seen in harmonically confined BECs with density-dependent gauge potentials \cite{Ishfaq_pre_2023}.
%%%%%%%%%%%%%%%%%%%%%%%% FIG 
%\FloatBarrier
%%%%%%%%%%%%%%%%%%%%%%%% 
The nucleation of the visible vortices results in a substantial increase in the 
system's angular momentum as shown in Fig. \ref{fig:lz_vs_om}(a) which displays 
the dependence of angular momentum on the trap rotation frequency with and 
without density-dependent gauge potentials. 
Fig. \ref{fig:lz_vs_om}(a) shows that for any value of $\tilde{\mathit{C}}$ the angular momentum increases linearly with $\Omega_{0}$ provided $\Omega_{0} < \Omega_{\text{cr}}$ and finally shows a jump at $\Omega_{\text{cr}}$. The linear progression in the angular momentum is due to an increase in the number of hidden vortices along the central barrier as shown in Fig. \ref{fig:om_cr_den}. It is worth mentioning that hidden vortices are associated with significant angular momentum. For $\Omega_{0}$ exceeding $\Omega_{\text{cr}}$, the angular momentum increases exponentially resulting in an increased number of visible vortices in the density profiles. Moreover, in Fig. \ref{fig:lz_vs_om}(a) one can see that at a fixed value of $\Omega_{0}$, the values of angular momenta are different and depend on the nature of nonlinear rotation, $\tilde{\mathit{C}}$. This can be explained from the corresponding phase profiles in Fig. \ref{fig:om_cr_den} whereby it is clear that the nonlinear rotation adjusts the number and/or the separation between the hidden vortices. The nonlinear rotation is even seen to affect the angular momentum in a BEC by fixing the positions of the visible vortices. 
%%%%%%%%%%%%%%%%%%%%%%%% FIG 
%\FloatBarrier
%%%%%%%%%%%%%%%%%%%%%%%%%%%%%%%%
In Fig. \ref{fig:lz_vs_om}(b) that shows  the vortex-ground states at the critical frequencies, it is evident that the visible vortex pair corresponding  to $\tilde{\mathit{C}}>0$ are closer to the central barrier compared to the ones with $\tilde{\mathit{C}}\le 0$. Since the angular momentum associated with a singly quantized defect varies with its position $r$ with respect to the centre as $l_{z} = (1-r^2/R^2)$ where $R$ is the radius of the BEC. Consequently, the angular momentum in the rotating BECs with density-dependent gauge potential follows $\braket{L_{z}}(\tilde{\mathit{C}} >0) > \braket{L_{z}}(\tilde{\mathit{C}} =0)> \braket{L_{z}}(\tilde{\mathit{C}} <0)$ for a given value of $\Omega_{0}$.
The dependence of vortex positions and the average angular momentum on the nature of nonlinear 
rotation in BECs with density-dependent gauge potential suggests a possible deviation 
from the standard Feynman's rule, ~$\braket{L_{z}}= N_{t}/2$. 
\par
\begin{figure}[!htb]
\centering
\includegraphics[width=0.45\textwidth]{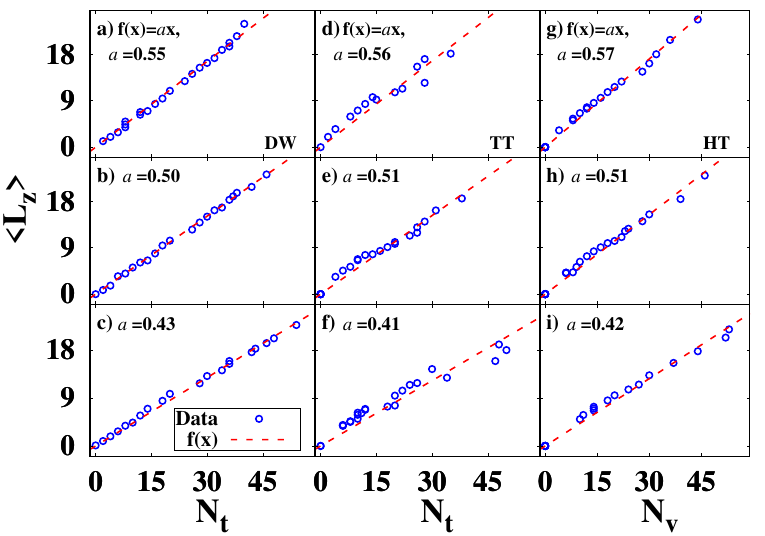}
\caption{ Linear fitting of $\braket{L_{z}}$ versus $N_{t}$ and $N_{v}$ 
corresponding to the vortex-ground states in a BEC with  $\tilde{g} = 420$ 
and trapped in (a-c) DW, (d-f) toroidal (TT) and (g-i) harmonic traps (HT) 
respectively. The nonlinear rotation strength, $\tilde{\mathit{C}} = 5 
~(\text{a,d,g}), 0 ~(\text{b,e,h}), \text{and}~  
 -10 ~(\text{c,f,i})$ respectively.}\label{fig:lz_vs_nt}
\end{figure}

\begin{figure}[!htb]
\centering
\includegraphics[width=0.45\textwidth]{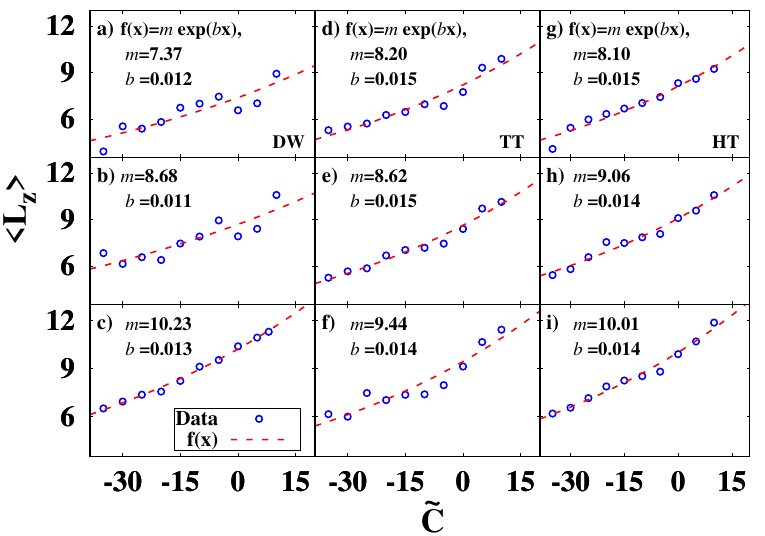}
\caption{ Exponential fitting of $\braket{L_{z}}$ versus $\tilde{\mathit{C}}$ 
corresponding to the vortex-ground states in a BEC with  $\tilde{g} = 420$ and 
trapped in (a-c) DW, (d-f) toroidal (TT) and (g-i) harmonic traps (HT) respectively. 
The number of vortices, $N_{t} = 14~\text{(a,d,g)}, 16~\text{(b,e,h)} \text{and}~ 
18~\text{(c,f,i)} $ respectively.}\label{fig:lz_vs_c}
\end{figure}

 
For a BEC in the absence of the density-dependent gauge potentials,  
$\tilde{\mathit{C}}=0$, it is known that Feynman's rule is satisfied 
with the inclusion of the hidden vortices. 
%%%%%%%%%%%%%%%%%%%%%%%% FIG 
\begin{figure}[!htb]
\centering
\includegraphics[width=0.45\textwidth]{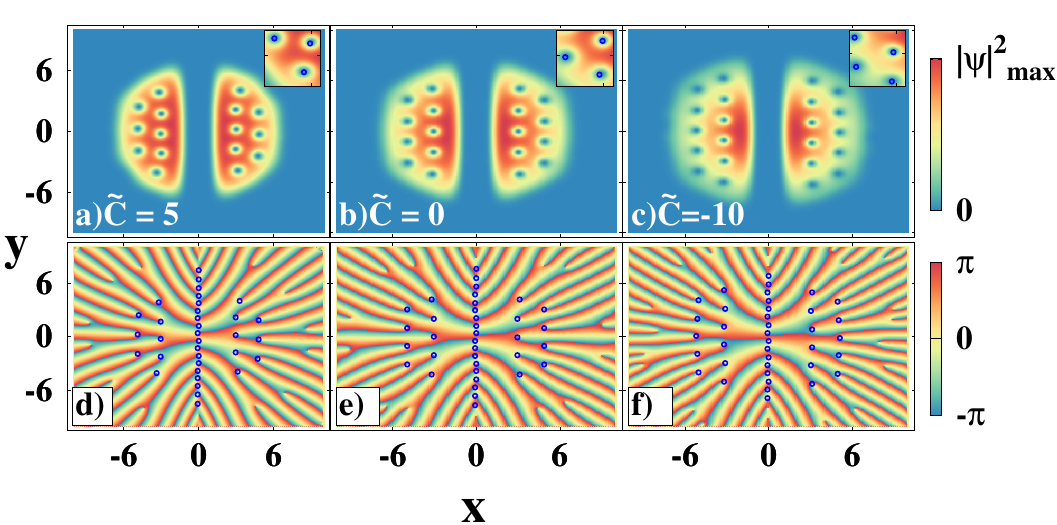}
\includegraphics[width=0.45\textwidth]{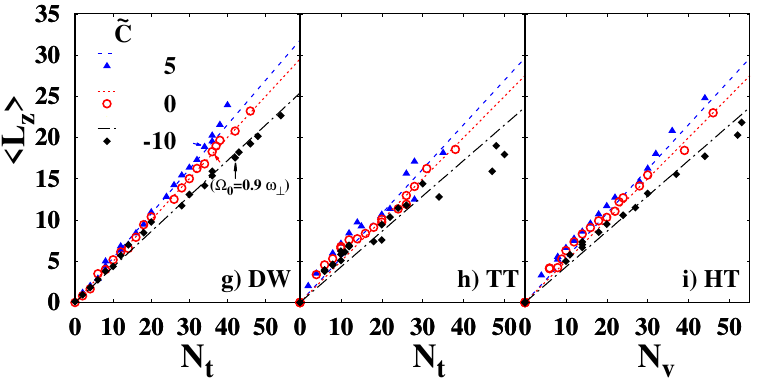}
\caption{ Density (a-c) and the corresponding phase profiles (d-f) of the vortex-ground states in a BEC with  $\tilde{g} = 420$ and rotated with $\Omega_{0}=0.9 \omega_{\perp}$ in a DW potential for different strengths of nonlinear rotation, $\tilde{\mathit{C}}$. The corresponding insets show vortex pattern in the region  $\sim [-5.4,-2.5]\times[-5.4,-1.4]$. (g-i)  Plot of Feynman's rule (Eq. \eqref{eq:Feynman_ddbec}) with lines and the numerical data with plot-markers for different strengths of nonlinear rotation, $\tilde{\mathit{C}}$ in (g) double well (DW), (h) toroidal (TT) and (i) harmonically (HT) trapped BECs. The data points identified by the arrows in (g) correspond to $\Omega_{0}=0.9\omega_{\perp}$ }\label{fig:Feynman_rule}
\end{figure}
%\FloatBarrier
As shown in Fig. \ref{fig:Feynman_rule}(b) the ground state of a BEC rotating in a 
DW potential at $\Omega_{0}= 0.9 \omega_{\perp}$ consists of a pair of vortex-lattices possessing hexagonal symmetry. However, the number of visible vortices, $N_{v}=18$ 
is much less to satisfy Feynman's rule, $\braket{L_{z}}=N_{v}/2$. By including the 
hidden vortices, $N_{h} = 18$ for which the average angular momentum, 
$\braket{L_{z}}\sim 18$, Feynman's rule, $\braket{L_{z}}=(N_{v}+N_{h})/2=N_{t}/2$ 
is satisfied as confirmed by the data point marked with a red arrow in 
Fig. \ref{fig:Feynman_rule}(g). The numerical data in the case of 
$\tilde{\mathit{C}}=0$ grazes along the respective theoretical curve  
and Feynman's rule in its standard form is satisfied for any number of vortices 
as shown in \ref{fig:Feynman_rule}(g). Feynman's rule for a BEC confined in a 
DW potential is derived in Ref. \cite{Mithun_pra_2014}. However, the spatial 
dependence of the rotation in BECs with density-dependent gauge potentials 
invalidates the formalism.  In the presence of the density-dependent gauge 
potentials where $\tilde{\mathit{C}}\ne 0$, Feynman's rule is more or less satisfied 
for a small number of vortices within the BEC. However, the system shows significant 
deviations from the standard Feynman's rule at large vortex numbers. These deviations 
can be attributed to the non-Abrikosov nature (lack of hexagonal symmetry) 
of the  visible vortex-lattices in BECs with density-dependent gauge 
potentials \cite{Edmonds_pra_2020,Edmonds_pra_2021,Ishfaq_pre_2023}, as confirmed in the insets of Figs. \ref{fig:Feynman_rule}(a,c). 
In contrast, the visible vortex-lattice in the case of $\tilde{\mathit{C}} = 0$ 
exhibits hexagonal symmetry (Abrikosov lattice, equilateral triangles) as shown in the inset of Fig. \ref{fig:Feynman_rule}(b). 
The nature of the nonlinear rotation, $\tilde{\mathit{C}}$ determines the 
amount of deviation. As indicated by the density profiles depicted in 
Fig. \ref{fig:Feynman_rule}, for the scenario where $\tilde{\mathit{C}}>0$, 
the visible vortices exhibit a closer proximity to the barrier such that 
$\braket{L_{z}}>N_{t}/2$. This positioning of the vortices causes the numerical 
data points to lie above the theoretical curve for $\tilde{\mathit{C}}=0$. 
On the other hand, for $\tilde{\mathit{C}}<0$, there is an increased presence 
of the vortices towards the outer periphery of the BEC, resulting in 
$\braket{L_{z}}<N_{t}/2$. 
The variation of average angular momentum $\braket{L_{z}}$ with the total number of 
vortices, $N_{t}$ in BECs with density-dependent gauge potentials can be explained by
modifying the standard form of Feynman's rule as
\begin{equation}\label{eq:Feynman_ddbec}
    \braket{L_{z}} = \frac{N_{t}}{2}
    ~\text{exp}\bigg(\frac{\tilde{\mathit{C}}\rho_{0}}{2}\bigg),
\end{equation}
where $\rho_{0}$ denotes the peak density of the BEC.
Equation \eqref{eq:Feynman_ddbec} represents an empirical relationship that 
explains the role of nonlinear rotation in determining the angular momentum 
characteristics in BECs with density-dependent gauge potentials. 
This equation highlights how the average angular momentum, given a certain number of 
vortices, can either increase or decrease depending on the nature of the nonlinear 
rotation at the maximum density, $\tilde{\mathit{C}}\rho_{0}$ and is illustrated in 
Fig. \ref{fig:Feynman_rule}(g-i) for respective confinements. Notably, the average angular momentum exhibits an 
exponential dependence on the strength of the nonlinear rotation, $\tilde{\mathit{C}}$ 
for a fixed value of peak density, $\rho_{0}$. In the special case where 
$\tilde{\mathit{C}}=0$, one recovers the standard form of Feynman's rule which 
typically applies to the conventional BECs with only rigid-body rotation. 
The results presented in this section find further support through the application 
of curve fitting analysis. As illustrated in Figures \ref{fig:Feynman_rule} and 
\ref{fig:lz_vs_nt}, the linear regression of average angular momentum ($\braket{L_{z}}$) against the total number of vortices ($N_{t}$) is depicted for fixed values of 
$\tilde{\mathit{C}}$. 
%%%%%%%%%%%%%%%%%%%%%%%% FIG 
%\FloatBarrier
%%%%%%%%%%%%%%%%%%%%%%%%%%%%
It is found that $\braket{L_{z}}$ varies linearly with the total 
number of vortices in the rotating BEC with the slope of the fitting line
depending on the nature of the nonlinear rotation arising from the density-dependent 
gauge potentials. We have disregarded the intercepts as they don't provide any 
valuable information, except for the presence of minimal rotation due to the 
density-dependent gauge potentials. The slope of the fitting line increases with the 
strength of $\tilde{\mathit{C}}$. For a given slope ($a$), the nonlinear rotation, 
$\tilde{\mathit{C}} = (2/\rho_{0})\text{ln}\left(2a\right)$. The specific values 
of $a$ in Fig. \ref{fig:lz_vs_nt}(a-c) lead to corresponding $\tilde{\mathit{C}}$ 
values of $\text{a})~ 6.35, \text{b})~ 0, ~\text{and}~ \text{c)}~ -10.05$ for a BEC 
confined in a DW potential. On the other hand, in the case of harmonically confined 
BECs, where only visible vortices are present, the resulting $\tilde{\mathit{C}}$ 
values are approximately $\text{g})~ 8.73, \text{h})~ 1.32, ~\text{and}~ \text{i})~ -11.62$ 
respectively. These derived values of $\tilde{\mathit{C}}$ closely align with the 
parameter values employed in numerical simulations, thus providing a robust validation 
of Eq. \eqref{eq:Feynman_ddbec}. We have further corroborated the aforementioned 
discussion through an examination of a BEC confined in a TT given in Eq. \eqref{eq:tt_pot}. 
The results as mentioned in Fig. \ref{fig:lz_vs_nt}(d-f) agree well with the 
empirical relation Eq. \eqref{eq:Feynman_ddbec}.
On similar lines, Fig. \ref{fig:lz_vs_c}  illustrates the exponential correlation between 
$\braket{L_{z}}$ and $\tilde{\mathit{C}}$. In this context, the fitted values 
$\textit{m}$ and $\textit{b}$ are directly related to the total number of vortices and the peak 
density of the BEC through the relationships, $\textit{m}=N_{t}/2$ and $\textit{b} = \rho_{0}/2$ 
respectively (Eq. \eqref{eq:Feynman_ddbec}). In Fig. \ref{fig:lz_vs_c}, we employed 
the BEC profiles characterized by different total number of vortices, specifically
$N_{t} = 14 ~\text{(a,d,g)}, 16 ~\text{(b,e,f)}, ~\text{and}~ 18 \text{(c,f,i)}$ 
respectively while the peak density, $\rho_{0}\sim 0.03$ in dimensionless units. 
The exponential fitting of $\braket{L_{z}}$ 
vs $\tilde{\mathit{C}}$ predicts the values for $N_{t}$ and $\rho_{0}$ around the 
numerically used values in each case. 
%%%%%%%%%%%%%%%%%%%%%%%% FIG 
%\FloatBarrier
%%%%%%%%%%%%%%%%%%%%%%%%
To illustrate, in Fig. \ref{fig:lz_vs_c}(a) 
depicting a DW-confined BEC, the fitting parameters $\textit{m}=7.37$ and $\textit{b}=0.012$ yield 
the estimated value of $N_{t} = 14$ and $\rho_{0}=0.024$ respectively. Remarkably, 
the estimated values in any case closely match with the numerically used parameters, 
underscoring the robustness of Eq. \eqref{eq:Feynman_ddbec} in predicting the vortex 
count and peak density. Consequently, this confirms the exponential dependence of 
$\braket{L_{z}}$ on $\tilde{\mathit{C}}$ in BECs with density-dependent gauge potentials.  
\par
\begin{figure}[!tb]
\centering
\includegraphics[width=0.45\textwidth]{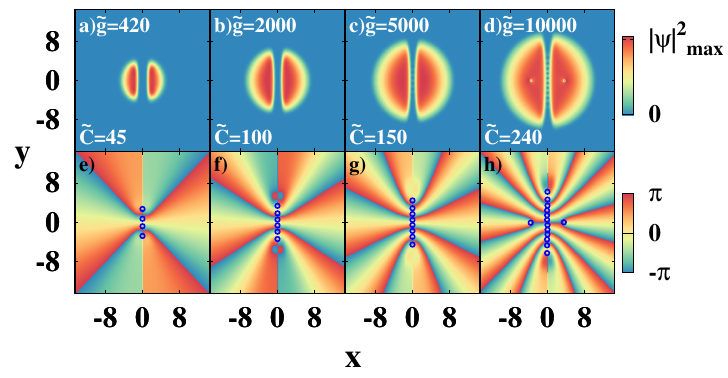}
\caption{ The nucleation of hidden and visible vortices 
in the ground states of a BEC corresponding to different 
two-body interactions, $\tilde{g}$ and nonlinear rotations, $\tilde{\mathit{C}}$. The BEC rotates in a static DW potential ($V_{0}=40$ and $\sigma = 0.5$) only via nonlinear rotation ($\Omega_{0}=0$).
}\label{fig:hid_to_vis_vor}
\end{figure}

\begin{figure}[!htb]
\centering
\includegraphics[width=0.45\textwidth]{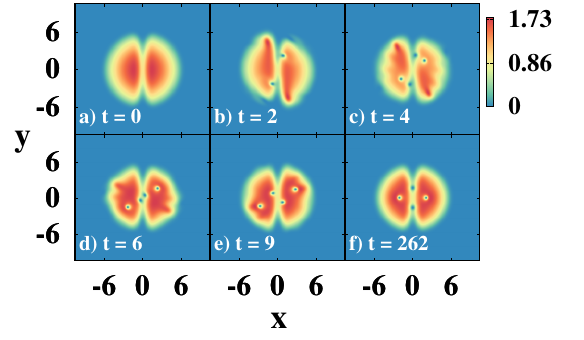}
\caption{Time development of the condensate density during vortex nucleation in a BEC with  $\tilde{g} =1000$ and solely rotated via nonlinear rotation ($\Omega_{0}=0$). The strength of the nonlinear rotation is maintained at $\tilde{\mathit{C}}=65$ while the confining potential is a DW of height $V_{0} = 10$ and width $\sigma=0.5$. 
}\label{fig:vis_vor_dy}
\end{figure}

 
Until now, we focused on a BEC with or without dynamical gauge potentials and 
subjected to external trap rotation. In general, it is found that visible vortices 
are nucleated only when the trap rotation exceeds the critical value. For trap 
rotations exceeding the critical value, vortex-lattices are formed. However, 
in BECs confined in complex potentials (DW and TT) hidden vortices  exist prior 
to the nucleation of the visible vortices at the critical trap rotation. The 
hidden vortices lie within the central barrier of the complex potential while 
the visible vortices form lattices. We now explore the possibility of vortex 
nucleation solely through the influence of dynamical gauge potentials in a BEC 
confined within an anisotropic and non-rotating complex potential. 
%%%%%%%%%%%%%%%%%%%%%%%%%%% 
\begin{figure}[!b]
\centering
\includegraphics[width=0.45\textwidth]{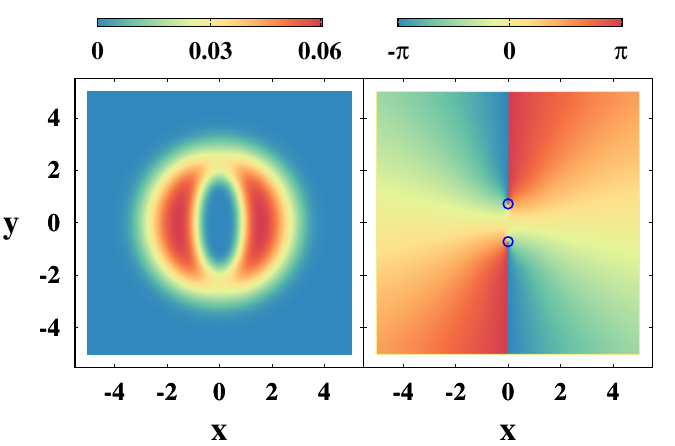}
\caption{ The nucleation of hidden vortices in the ground states of a toroidally confined  BEC corresponding to $\tilde{g} = 100$ and nonlinear rotation, $\tilde{\mathit{C}} = 23.5$. The BEC rotates in a static TT ($V_{0}=40$, $\beta = 2.5$ and $\sigma = 0.5$) solely through nonlinear rotation ($\Omega_{0}=0$).
}\label{fig:hid_vor_om_0_tt}
\end{figure}
%\FloatBarrier
%%%%%%%%%%%%%%%%%%%%%%%%%%%
Since the vortex nucleation solely due to the nonlinear rotation ($\Omega_{0}=0$) 
is anticipated for a BEC with large number of atoms 
\cite{Butera_jpb_2016,Butera_pt_2017}. 
When considering a harmonically confined BEC in the presence of density-dependent 
gauge potentials, it is found that the vortices do not nucleate solely due to the 
nonlinear rotation \cite{Ishfaq_pre_2023}. The same is found true for a BEC with 
large number of atoms and confined in a slightly anisotropic toroidal trap with 
small central barrier heights. For small BECs trapped in toroidal traps with large 
central barrier heights and anisotropy, only hidden vortices are possible as shown 
in Fig. \ref{fig:hid_vor_om_0_tt}. However, in the case of DW-confined BECs, 
Fig. \ref{fig:hid_to_vis_vor} illustrates numerically simulated ground states 
of a BEC for different values of nonlinear interaction strengths. For a given 
DW-potential, the BEC exhibits hidden vortices exclusively when the two-body 
interaction strengths are relatively low. The number of hidden vortices 
increases with the strength of the two-body interactions and hence the 
nonlinear rotation. 
%%%%%%%%%%%%%%%%%%%%%%%%%%% 
%\FloatBarrier
%%%%%%%%%%%%%%%%%%%%%%%%%%% 
The hidden vortices along the central barrier of the DW potential start becoming visible when the BEC density around the central barrier increases. At much larger interaction strengths, when the overlap along the central barrier is complete, visible vortices appear along the low-density central barrier region and within the bulk of the BEC. The density-dependent gauge potentials induce the nucleation of only a small number of vortices, even with significantly large two-body interactions, on either side of the barrier. Consequently we did not observe any vortex-lattice formation. 
The nucleation of visible vortices via the nonlinear rotation alone at large interaction strengths hints at the possibility of vortex nucleation even for smaller two-body interaction strengths provided the height of the barrier is lowered. 
In this direction, Fig. \ref{fig:vis_vor_dy} shows the dynamics of vortex nucleation due to the nonlinear rotation in a DW-confined BEC with $\tilde{g} = 1000, V_{0} = 10, ~\text{and}~ \tilde{C} = 65$. 
%%%%%%%%%%%%%%%%%%%%%%%%%%%
%\FloatBarrier
%%%%%%%%%%%%%%%%%%%%%%%%%%%
The initial deformation of the BEC leads to the creation of a vortex pair at the inner surface of the barrier. The vortices then subsequently move into the bulk of the condensate by following the motion of the rotating BEC. 
The BECs with opposite signs of the nonlinear rotation rotate in opposite directions but result in the same number of the vortices. It is worth mentioning that the vortices still enter the BEC through the low-curvature region \cite{Tsubota_pra_2003}. However, the shape oscillations induced in the BEC are aperiodic because of the nonlinear rotation due to the density-dependent gauge potentials \cite{Ishfaq_pre_2023}. Moreover, it is found that in addition to the interaction strength values, the number of vortices nucleated by the nonlinear rotation is also limited by the shape and size of the potential. 
\par
\begin{figure}[!htb]
\centering
\includegraphics[width=0.45\textwidth]{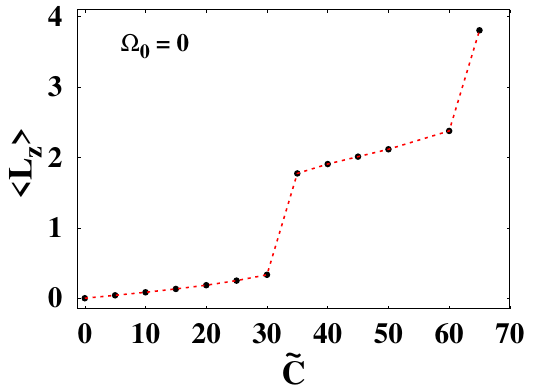}
\caption{Variation of average angular momentum, $\braket{L_{z}}$ with the strength of nonlinear rotation, $\tilde{\mathit{C}}$ in a BEC with  $\tilde{g} =1000$ and no external trap rotation ($\Omega_{0}=0$) in a DW potential of height $V_{0} = 10$ and width $\sigma=0.5$.}\label{fig:lz_vs_nr_og_0}
\end{figure}

In the above described settings, Fig. \ref{fig:lz_vs_nr_og_0} shows the variation of 
average angular momentum, $\braket{L_{z}}$ with the strength of the density-dependent 
gauge potential. It is evident that angular momentum increases linearly with 
$\tilde{\mathit{C}}$ before the nucleation of a new pair of vortices. The angular 
momentum undergoes abrupt changes with the addition of each 
successive pair of vortices.
Additionally, it is observed that for the vortices nucleated solely by means of 
density-dependent gauge potentials, the average angular momentum is 
proportional to the total number of vortices, \textit{i.e.}, $\braket{L_{z}} \sim N_{t}$.
%%%%%%%%%%%%%%%%%%%%%%%%%%% FIG
%\FloatBarrier
%%%%%%%%%%%%%%%%%%%%%%%%%%%%%%%%%
%%%%%%%%%%%%%%%%% Conclusion
\section{Conclusions}\label{sec:concl}
In this work we have considered a BEC in the presence of density-dependent gauge potential and confined either in a harmonic trap (HT), a DW-potential or a toroidal trap (TT). We have examined the validity of the standard Feynman’s rule in BECs featuring the density-dependent gauge potentials. Because of the nonlinear rotation arising due to the density-dependent potentials, these BECs do not follow the standard form of Feynman’s rule. The angular momentum for a given number of vortices also depends on the nature of the nonlinear rotation and the peak density of the BEC. We presented an empirical relation as modified Feynman’s rule that relates the average angular momentum and the number of vortices in BECs with density-dependent potentials. The average angular momentum alters in an exponential manner with the strength of the density-dependent gauge potentials. The agreement between the modified Feynman's rule and the numerical data is well established through fitting analysis.
\par 
Moreover, within the context of a harmonically confined BEC, it is demonstrated that the mere presence of density-dependent gauge potentials alone does not lead to vortex nucleation, even when the mean-field interactions are considerable \cite{Ishfaq_pre_2023}. However, in case of complex potentials, density-dependent gauge potentials alone can result in vortex nucleation. We observe only hidden vortex formation due to the density-dependent gauge potentials alone in BECs confined in a highly anisotropic TT. Similarly, aided by the anisotropy in the DW confinements, our findings reveal that nucleation of visible vortices in BECs is even achievable solely through density-dependent gauge potentials. This phenomenon is particularly notable in DW confinements characterized by low central barrier heights. For large barrier heights, only hidden vortices are obtained. The number of nucleated vortices is not only determined by the interaction parameters but are also limited by the specifications of the DW-potential. Further the average angular momentum varies proportionately with the number of vortices in case the BEC is rotated solely by means of density-dependent gauge potentials, $\braket{L_{z} }\sim N_{t}$. We believe that the recent two experiments \cite{frolian_nat_2022, yao_nat_2022} involving coupling of density-dependent gauge fields with Raman coupled BEC in a 1D can be generalized to 2D, as considered in the manuscript, in near future.
\section{Acknowledgements}
BD thanks Science and Engineering Research
Board, Government of India for funding through research 
project CRG/2020/003787. I A Bhat acknowledges CSIR, 
Government of India, for funding via CSIR 
Research Associateship (09/137(0627)/2020 EMR-I). 
%%%%%%%%%%%%%%%%%%%%%%%%%%%%%%%%%%%%%%%%%%%%
%%%%%%%%%%%%%%%%%%%%%%%%%%%%%%%%%%%%%%%%%%%%
\let\itshape\upshape\normalem
\bibliographystyle{apsrev}
\bibliography{references_hid_ddbec}

\end{document}